\documentclass[english,aps, twocolumn]{revtex4}
\usepackage[T1]{fontenc}
\usepackage[latin9]{inputenc}
\usepackage{amsmath}
\usepackage{amssymb}
\usepackage{graphicx}
\usepackage{esint}

\makeatletter

\providecommand{\tabularnewline}{\\}

\@ifundefined{textcolor}{}
{%
 \definecolor{BLACK}{gray}{0}
 \definecolor{WHITE}{gray}{1}
 \definecolor{RED}{rgb}{1,0,0}
 \definecolor{GREEN}{rgb}{0,1,0}
 \definecolor{BLUE}{rgb}{0,0,1}
 \definecolor{CYAN}{cmyk}{1,0,0,0}
 \definecolor{MAGENTA}{cmyk}{0,1,0,0}
 \definecolor{YELLOW}{cmyk}{0,0,1,0}
 }

\@ifundefined{date}{}{\date{July 11th, 2015}}
\makeatother

\usepackage{babel}
\begin{document}

\title{Low emittance pion beams generation from bright photons and relativistic protons}

\author{L. Serafini, C. Curatolo and V. Petrillo}

\affiliation{INFN-Milan and Università degli Studi di Milano, via Celoria 16, 20133 Milano, Italy}

\begin{abstract}
Present availability of high brilliance photon beams as those produced by X-ray Free Electron Lasers in combination with intense TeV proton beams typical of the Large Hadron Collider makes it possible to conceive the generation of pion beams via photo-production in a highly relativistic Lorentz boosted frame: the main advantage is the low emittance attainable and a TeV-class energy for the generated pions, that may be an interesting option for the production of low emittance muon and neutrino beams. We will describe the kinematics of the two classes of dominant events, i.e. the pion photo-production and the electron/positron pair production, neglecting other small cross-section possible events like Compton and muon pair production. Based on the phase space distributions of the pion and muon beams we will analyze the pion beam brightness achievable in three examples, based on advanced high efficiency high repetition rate FELs coupled to $LHC$ or Future Circular Collider ($FCC$) proton beams, together with the study of a possible small scale demonstrator based on a Compton Source coupled to a Super Proton Synchrotron ($SPS$) proton beam.
\end{abstract}

\maketitle

\section{Introduction}
One of the main challenges of present muon collider design studies is the capture/cooling stage of muons after generation by intense GeV-class proton beams impinging on solid targets: this mechanism produces pions further decaying into muons and neutrinos. As extensively analyzed in Ref. \cite{Ankenbrandt,Alsharo}, the large emittance of the generated pion beams, which is mapped into the muon beam, is mainly given by the mm-size beam source at the target (i.e. the proton beam focal spot size) and by Coulomb scattering of protons and pions propagating through the target itself, inducing large transverse momenta which in turns dilute the phase space area. Recently the option to use hundreds of MeVs photons to make muon pairs via photo-production in a solid target was analyzed in Ref. \cite{mcdonald}, with the aim of getting a smaller source spot size, avoiding as well the proton beam scattering effects. This scheme is promising in terms of low emittance of the muon beams at the source, but unfortunately it turns out that the low energy of the generated muons and the transport process into the high field solenoid system  induces an irreversible emittance growth. Therefore the small source emittance of muons produced in the Bethe-Heitler process does not compensate its very low efficiency \cite{mcdonald2}.

We propose in this paper a different approach, still based on photon beams, enabled by the present availability of outstanding TeV-class high intensity proton beams and ultra-high brilliance X-ray photon beams obtained by advanced X-ray Free Electron Lasers: although these two classes of accelerators have been developed in the last decades adopting completely different technologies and aiming at totally separated research domains, they both represent historical milestones in the development of large scale accelerator based research infrastructures for fundamental and applied research. In other words accelerators like $LHC$ and $LCLS/XFEL$ \cite{lhc, lcls, xfel} have become major tools of discovery for the ongoing research and will remain main actors for the decades to come. Their combined capability of producing ultra-high phase space density particle beams is the base of our strategy for generating low emittance pion, muon and neutrino beams, using collisions between two counter-propagating beams of highly relativistic protons and ultra-high intensity photons. The extremely high luminosity achievable by such a collider ($10^{38}$ cm$^{-2}$s$^{-1}$) can compensate for the low efficiency of the pion photo-production which has a total cross section of $\simeq 220 $ $\mu$barn with $300$ MeV photons, much smaller than GeV-proton based pion production ($\simeq 20$ mbarn).

There are two crucial aspects in such a collision scheme. The first is the much higher energy of the X-ray photons observed by the proton in its own rest frame: this enables pion photo-production above the threshold with maximum efficiency, despite the keV energy of the colliding photon. The second deals with the proton carrying almost the total momentum of the system, which makes it the source of highly Lorentz boosted secondary beams collimated within a narrow forward angle of the same order of the proton beam diffraction angle given by its transverse emittance (tens of $\mu$rads). In this sense the mechanism for pion production described in the following represents sort of a relativistic pion photo-cathode. A similar approach has been proposed and discussed in Ref. \cite{dadi} although based on a multi-photon production of pion and muon pairs, which is a much lower efficiency process.

We considered two examples of TeV proton beams: an upgraded $LHC$ beam carrying up to $10^{12}$ protons per bunch at $7$ TeV energy and an $FCC$ expected beam at the same bunch intensity with $50$ TeV energy, both with a repetition rate up to $40$ MHz.

Photon beams at energies ranging between $3$ and $20$ keV and fluxes larger than $10^{13}$ photons per pulse at MHz repetition rate can be obtained with superconductive $CW$ Linac based XFELs, in saturation regime for the energies lower than few keV and in the tapering mode for larger energies. The brilliance of XFEL photon beams coupled to X-ray optics technology \cite{yumoto, jiao} allows focusing the FEL photon beam down to spot sizes $\simeq10$ $\mu$m comparable to $LHC$ proton beam at the interaction point (IP), in order to maximize the luminosity of proton-photon collider. In table \ref{tab:FEL-performance}, data taken from Ref. \cite{Pelle,key-1} show the status of art of the FEL simulations
in these regimes.
\begin{table}
	\begin{tabular}{|c|c|c|c|c|}
		\hline 
		Ref. & Mode & $h\nu$ & e-beam  & Photon \#  \tabularnewline
		\hline 
		\hline 
		\cite{Pelle} & Tapering & $8.25$ keV & $160$ pC & $3\cdot 10^{13}$\tabularnewline
		\hline 
		\cite{key-1} & Saturation & $<1.65$ keV & $1$ nC & $>10^{13}$\tabularnewline
		\hline 
		\cite{key-1} & Tapering & $<3$ keV & $>0.1$ nC & $10^{13}\div10^{14}$\tabularnewline
		\hline 
		\cite{key-1} & Tapering & $<2$ keV & $1$ nC & $>10^{14}$\tabularnewline
		\hline 
		\cite{key-1} & Tapering & $20$ KeV &  $>0.2$ nC & $2\cdot 10^{13}$\tabularnewline
		\hline 
	\end{tabular}
	\caption{\label{tab:FEL-performance}FEL performance}
\end{table}
Fluxes close or larger than $10^{13}$ photons per pulse are achieved for energies smaller than $1.65$ keV in the saturated mode. The tapering allows to increase the flux by one order of magnitude while enhancing the energy. By applying the FEL scaling laws to the data reported in Ref. \cite{key-1} for the case of hundreds femtoseconds long electron bunches allows to envision the possibility of exceeding $10^{13}$ photons at $20$ keV.

\section{Kinematics of pion/muon photo-production in a highly Lorentz boosted frame}

Let us consider the collision between a proton and a counter-propagating photon of energy respectively $E_{pr}$ and $E_{ph}$ in the laboratory frame. We are interested in the study of the reaction \begin{equation*}
p + h\nu \rightarrow \pi^{+} + n
\end{equation*}
We set $c=1$ and denote with $*$ the particles' momenta and energies in their center of mass reference frame. 

The energy $E_{ph}'$ of the colliding photon in the proton rest frame is given by the formula
\begin{equation}
 E_{ph}'=\gamma_{pr}E_{ph}(1-\underline{\beta}_{pr}\cdot\underline{e}_{k})
\end{equation}
 where $\underline{\beta}_{pr}$ is the velocity of the proton, $\underline{e}_{k}$ is the direction of propagation of the photon, $\gamma_{pr}=E_{pr}/M_{pr}$ and $M_{pr}=938$ MeV/c$^{2}$. 
For an ultra-relativistic proton colliding head-on with a photon, the formula simplifies in $E_{ph}'\simeq 2\gamma_{pr}E_{ph}$. As an example, if $E_{pr}=7$ TeV and $E_{ph}=20$ keV we have $E_{ph}'=298$ MeV.

The center of mass $(cm_{1})$ of the pion-neutron system is characterized by
\begin{equation}\gamma_{cm_{1}}=\frac{E_{tot}^{lab}}{E_{cm_{1}}}\simeq\frac{E_{pr}+E_{ph}}{\sqrt{4E_{pr}E_{ph}+M_{pr}^{2}}}\end{equation}
where $E_{cm_{1}}$ is the invariant mass of the proton-photon system. Using the same example as above we find $\gamma_{cm_{1}}=5834$.
Therefore the modulus of the momentum of the emitted $\pi$ in $cm_{1}$ is
\begin{equation}p^{*}_{\pi}=\frac{\sqrt{E_{cm_{1}}^{4}+(M_{\pi}^{2}-M_{n}^{2})^{2}-2E_{cm_{1}}^{2}(M_{\pi}^{2}+M_{n}^{2})}}{2E_{cm_{1}}}\label{pion}\end{equation}
where $M_{\pi}=139.6$ MeV/c$^{2}$ and $M_{n}=939.565$ MeV/c$^{2}$.
The energy of the pion in $cm_{1}$ is
\begin{equation}
E^{*}_{\pi}=\sqrt{p^{*2}_{\pi}+M_{\pi}^{2}}
\end{equation}
and by applying the Lorentz transformations \cite{jackson} we obtain the momentum components of $\pi$ in the laboratory frame\\

$\left\{
\begin{aligned}
&p_{x\pi}=p^{*}_{\pi} \sin{\theta^{*}}\cos{\phi^{*}}\\
&p_{y\pi}=p^{*}_{\pi} \sin{\theta^{*}}\sin{\phi^{*}}\\
&p_{z\pi}=\gamma_{cm_{1}}\left(\sqrt{1-\frac{1}{\gamma_{cm_{1}}^{2}}}E^{*}_{\pi}+p^{*}_{\pi}\cos{\theta^{*}}\right)
\end{aligned}
\right.$\vspace{0.5cm}\\
with $\theta^{*}$ and $\phi^{*}$ angles in $cm_{1}$.
The angle $\theta_{\pi}$ in the lab frame is
\begin{equation}\theta_{\pi}=\arctan\left(\frac{\sqrt{p_{x\pi}^{2}+p_{y\pi}^{2}}}{p_{z\pi}}\right)\end{equation}
considering $\theta_{\pi}=0$ the direction of the proton's propagation and
\begin{equation}\gamma_{\pi}=\frac{\sqrt{p_{x\pi}^2+p_{y\pi}^2+p_{z\pi}^2+M_{\pi}^2}}{M_{\pi}}\end{equation}
The other reaction product, the neutron, is treated with the same approach as the pion and it is characterized by $p^{*}_{n}=p^{*}_{\pi}$ and $E^{*}_{n}=\sqrt{p^{*2}_{n}+M_{n}^{2}}$.\\

For a pion at rest in the laboratory frame the mean life-time is $\tau_{\pi}=2.6\cdot 10^{-8}$ s to decay into a muon and a neutrino according to the following reaction: 
\begin{equation*}
	\pi^{+}\rightarrow\mu^{+}+\nu_{\mu}
\end{equation*}
In our case the pion mean life-time is $\gamma_{\pi} \cdot \tau_{\pi}$, which implies pions propagating over long distances (hundred meters to several kms). In the following we will populate the phase space volumes of the generated pion and muon beams neglecting effects coming from this long range pion propagation: the reason for this assumption is to focus the present analysis on the secondary beam generation process and leaving to a future work the study about matching the generated beams into a further storage/acceleration stage. \\

The muon momentum in its center of mass $(cm_{2})$ is
\begin{equation}p^{*}_{\mu}=\frac{M_{\pi}^{2}-M_{\mu}^{2}}{2M_{\pi}}\label{muon}\end{equation}
where $M_{\mu}=105.65$ MeV/c$^{2}$.
The energy of the muon in $cm_{2}$ is $E^{*}_{\mu}=\sqrt{p^{*2}_{\mu}+M_{\mu}^{2}}$ and the momentum components are\\

$\left\{
\begin{aligned}
&p_{x\mu}^{*}=p^{*}_{\mu} \sin{\theta^{*}}\cos{\phi^{*}}\\
&p_{y\mu}^{*}=p^{*}_{\mu} \sin{\theta^{*}}\sin{\phi^{*}}\\
&p_{z\mu}^{*}=p^{*}_{\mu} \cos{\theta^{*}}
\end{aligned}
\right.$\vspace{0.5cm}\\
with $\theta^{*}$ and $\phi^{*}$ angles in $cm_{2}$.

Since the pion decays while moving in the direction given by the components of its momentum, in order to define the components of the muon momentum in the laboratory frame, we use the Lorentz transformations for a boost in a generic direction with $\gamma_{cm_{2}}=\gamma_{\pi}$ as follows:\vspace{.5cm}

$\left\{
\begin{aligned}
&p_{x\mu}=\frac{E^{*}_{\mu}p_{x\pi}}{M_{\pi}}+p_{x\mu}^{*}\left(1+\frac{p_{x\pi}^{2}}{(\gamma_{\pi}+1)M_{\pi}^{2}}\right)\\
&+p_{y\mu}^{*}\left(\frac{p_{x\pi}p_{y\pi}}{(\gamma_{\pi}+1)M_{\pi}^{2}}\right)+p_{z\mu}^{*}\left(\frac{p_{x\pi}p_{z\pi}}{(\gamma_{\pi}+1)M_{\pi}^{2}}\right)\\
&p_{y\mu}=\frac{E^{*}_{\mu}p_{y\pi}}{M_{\pi}}+p_{x\mu}^{*}\left(\frac{p_{x\pi}p_{y\pi}}{(\gamma_{\pi}+1)M_{\pi}^{2}}\right)\\
&+p_{y\mu}^{*}\left(1+\frac{p_{y\pi}^{2}}{(\gamma_{\pi}+1)M_{\pi}^{2}}\right)+p_{z\mu}^{*}\left(\frac{p_{y\pi}p_{z\pi}}{(\gamma_{\pi}+1)M_{\pi}^{2}}\right) \\
&p_{z\mu}=\frac{E^{*}_{\mu}p_{z\pi}}{M_{\pi}}+p_{x\mu}^{*}\left(\frac{p_{x\pi}p_{z\pi}}{(\gamma_{\pi}+1)M_{\pi}^{2}}\right)\\
&+p_{y\mu}^{*}\left(\frac{p_{y\pi}p_{z\pi}}{(\gamma_{\pi}+1)M_{\pi}^{2}}\right)+p_{z\mu}^{*}\left(1+\frac{p_{z\pi}^{2}}{(\gamma_{\pi}+1)M_{\pi}^{2}}\right)
\end{aligned}
\right.$\vspace{.5cm}\\

In the case of a pure longitudinal boost, $p^{*}_{\pi}$ and $p^{*}_{\mu}$ as given in eq.(\ref{pion}), (\ref{muon}) represent the maximum value of the pion and muon transverse momentum also in the laboratory frame because of the invariance of transverse momenta under Lorentz transformations. By applying once again the example mentioned above and considering the proton beam without emittance (we will refer to it as parallel $\parallel$ beam), we find $p_{x\pi}, p_{x\mu} \le 195$ MeV/c. This is very relevant in order to generate a low emittance secondary beam: as a matter of fact the  normalized transverse emittance at the interaction point of the pion beam is given by  $\epsilon_{xn\pi}={\sigma_{xpr}}\sqrt{\langle \overline{p}_{x\pi}^{2}\rangle}$ where we use the dimensionless transverse momentum defined as $\overline{p}_{x\pi}\equiv p_{x\pi}/M_{\pi} $ and the average is performed over the phase space distribution area. It is clearly shown that for proton beam spot size at IP $\sigma_{xpr} \le 10$ $\mu$m we can generate pion beams with normalized emittances of a few mm$\cdot$mrads, as discussed below. As already mentioned the challenge of preserve such a low emittance in the decay of a pion to generate the muon beam is a matter of a future work.

One of the advantages of this pion photo-cathode scheme lays in the production of highly relativistic muons that are long lived, in excess of tens of milliseconds, allowing to conceive a complex and multistaged beam manipulation scheme, aimed at preserving the emittance and more in general the $6D$ phase space volume.\\

We create an event-generator code to simulate the proton-photon beams interaction and we illustrate in the following the results obtain for different parameters: $E_{pr}=7$ TeV (case $LHC$) and $E_{ph}=20$ keV, $E_{pr}=50$ TeV (case $FCC$) and $E_{ph}=3$ keV and $E_{pr}=400$ GeV (case $SPS$) and $E_{ph}=350$ keV. 

The photon energy $E_{ph}$ is chosen in order to maximize the Lorentz invariant total cross section \cite{heitler} for pion photo-production on protons $\simeq 220$ $\mu$barn corresponding to $E'_{ph}=300$ MeV \cite{tsai, drechsel2}, a figure obtained by proton-gamma collision experiments where solid or liquid targets are hit by bremsstrahlung generated photons \cite{hilpert}. The cross sections of neutrons and deuterons have similar values \cite{kondo, darwish}. 

In our simulation the pions are distributed in a circle of $10$ $\mu$m radius and the $\theta$ angle is taken with respect to the mean direction of the proton beam propagation.
The FEL photon beam is described by a diffraction limited radiation beam with geometric emittance $\epsilon_{h\nu}=\lambda_{FEL}/4\pi$: its diffraction angle at the focal a plane is given by $\sigma'_{h\nu}=\epsilon_{h\nu}/\sigma_{xh\nu}$. Assuming an optimal space-time overlap between the two colliding beams at the IP, i.e. $\sigma_{xpr}=\sigma_{xh\nu}$, we find an outstanding diffraction behavior of the FEL beam, represented by $\sigma'_{h\nu} \le 1$ $\mu$rad in the energy range $12<h\nu<20$ keV.
The minimum and maximum value of the longitudinal momentum, the maximum of $p_{x}$ and the value of $\theta$ angle are reported in tables \ref{tab:valori}, \ref{tab:valorifcc} and \ref{tab:valoridim} for all the kind of particles involved in the reaction.\\
\begin{table}[htbp]
	\begin{tabular}{|c|c|c|c|c|}
		\hline  Particle & $pz_{min}$ & $pz_{max}$ & $px_{rms}$ & $\theta_{rms}$ \\ 
		\hline  Proton $(pr)$ & $\simeq 7000$ & $7000$ & $0.187$ &  $27$ \\ 
		\hline  Photon $(ph)$ & $20$ keV & $20$ keV & $20$ meV & $0.6$ \\ 
		\hline  Pion $(\pi)$ & $260$ & $2540$ & $0.38$ & $145$ \\ 		
		\hline  Neutron $(n)$ & $4450$ & $6740$ & $0.38$ & $27$ \\  
		\hline  Muon $(\mu)$ & $149$ & $2500$ & $0.38$ & $147$ \\  
		\hline  Neutrino $(\nu)$ & $0$ & $1080$ & $0.16$ & $\le 500$ \\ 
		\hline 
	\end{tabular}
		\caption{Particles properties (case $LHC$). Values of momenta in [GeV/c], angles in [$\mu$rad].}
		\label{tab:valori}
\end{table}
\begin{table}[htbp]
	\begin{tabular}{|c|c|c|c|c|}
		\hline  Particle & $pz_{min}$ & $pz_{max}$ & $px_{max}$ & $\theta_{rms}$ \\ 
		\hline  Proton $(pr)$ & $\simeq 50000$ & $50000$ & $\parallel beam$ & $\parallel beam$\\ 
		\hline  Photon $(ph)$ & $3$ keV & $3$ keV & $140$ meV & $7$ \\ 
		\hline  Pion $(\pi)$ & $1722$ & $19108$ & $0.21$ & $11$ \\ 		
		\hline  Neutron $(n)$ & $30891$ & $48277$ & $0.21$ & $1.45$ \\  
		\hline  Muon $(\mu)$ & $986$ & $19108$ & $0.22$ & $11.5$ \\  
		\hline  Neutrino $(\nu)$ & $0$ & $8162$ & $0.1$ &  $\le 500$  \\ 
		\hline 
	\end{tabular}
	\caption{Particles properties (case $FCC$). Values of momenta in [GeV/c], angles in [$\mu$rad].}
	\label{tab:valorifcc}
\end{table}
\begin{table}[htbp]
	\begin{tabular}{|c|c|c|c|c|}
		\hline  Particle & $pz_{min}$ & $pz_{max}$ & $px_{max}$ & $\theta_{rms}$ \\ 
		\hline  Proton $(pr)$ & $\simeq 400$ & $400$ & $\parallel beam$ & $\parallel beam$\\ 
		\hline  Photon $(ph)$ & $350$ keV & $350$ keV & $0.19$ keV & $540$ \\ 
		\hline  Pion $(\pi)$ & $14.9$ & $145$ & $0.195$ & $1200$ \\ 		
		\hline  Neutron $(n)$ & $254$ & $385$ & $0.195$ & $166$ \\  
		\hline  Muon $(\mu)$ & $8.5$ & $145$ & $0.2$ & $1300$ \\  
		\hline  Neutrino $(\nu)$ & $0$ & $62$ & $0.093$ &  $9$ mrad  \\ 
		\hline 
	\end{tabular}
	\caption{Particles properties (case $SPS$). Values of momenta in [GeV/c], angles in [$\mu$rad].}
	\label{tab:valoridim}
\end{table}\\
Two additional cases have been considered: $E_{pr}=7$ TeV and $E_{pr}=50$ TeV colliding with photons at $E_{ph}=12$ keV (see tables \ref{tab:valorialtri2} and \ref{tab:valorialtri}).
The decrease of $E'_{ph}$ at constant proton energy leads to an increase of the minimum and to a decrease of the maximum value of $pz$ for all the particles involved (compare table \ref{tab:valori} to \ref{tab:valorialtri2} and \ref{tab:valorifcc} to \ref{tab:valorialtri}); the minimum and maximum values of $pz$ coincide at the threshold value of the reaction which is $E'_{ph}=150$ MeV. Therefore at threshold we would produce two quasi-monochromatic beams of pions and neutrons, although with a quite low efficiency. One of the promising aspects of this scheme is the tunability of the FEL wavelength, which allows an easy adjustment of the photon energy in order to optimize the collision to the desired performances.

\begin{figure}[htbp]
	\includegraphics [scale=0.21] {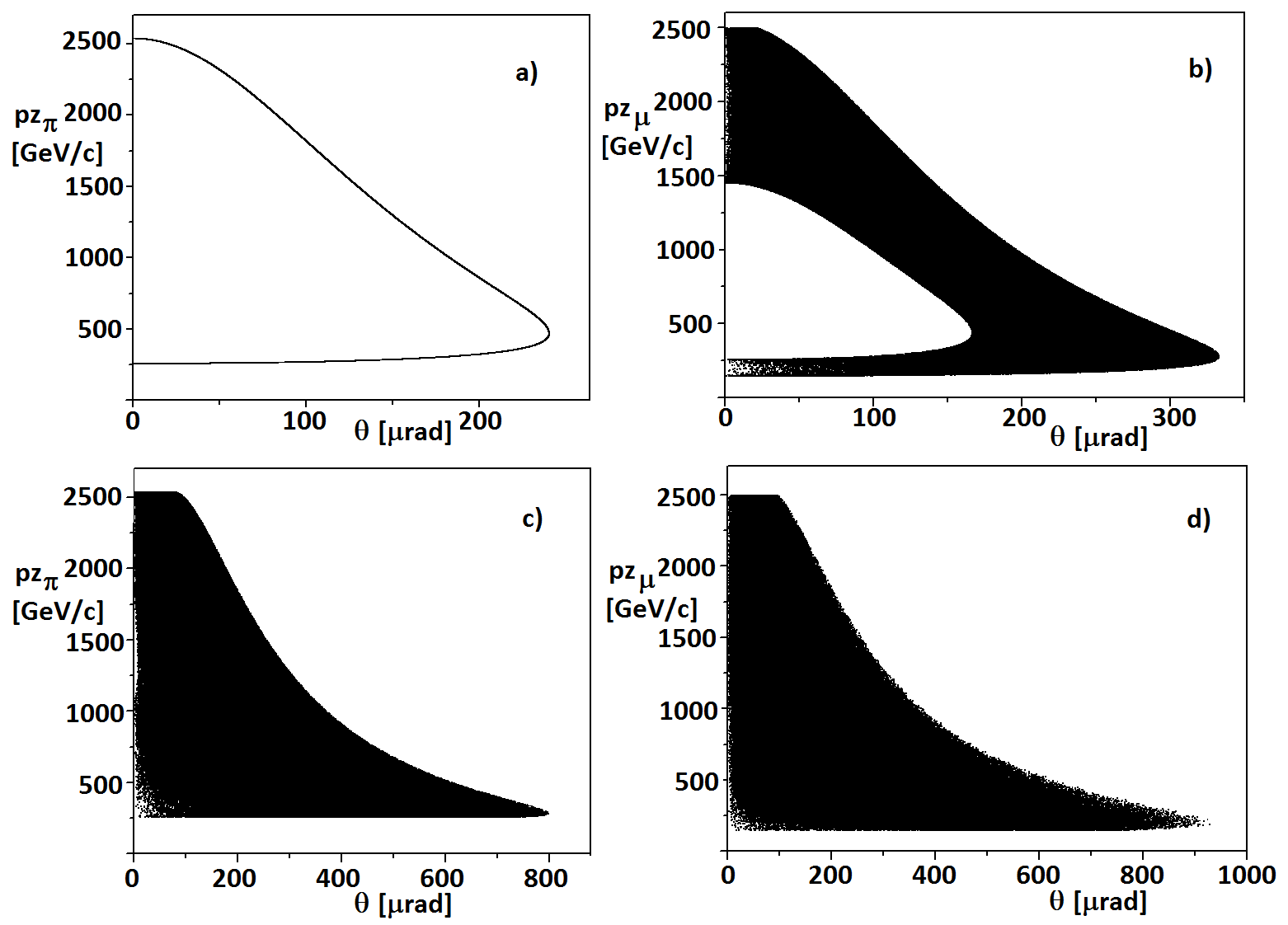}
	\caption{Pion and muon longitudinal momentum [GeV/c] as a function of $\theta$ [$\mu$rad], without (top) and with (bottom) transverse emittance of the incoming proton beam (case $LHC$).}\label{pimu}
\end{figure}

\begin{figure}[htbp]
	\includegraphics [scale=0.21] {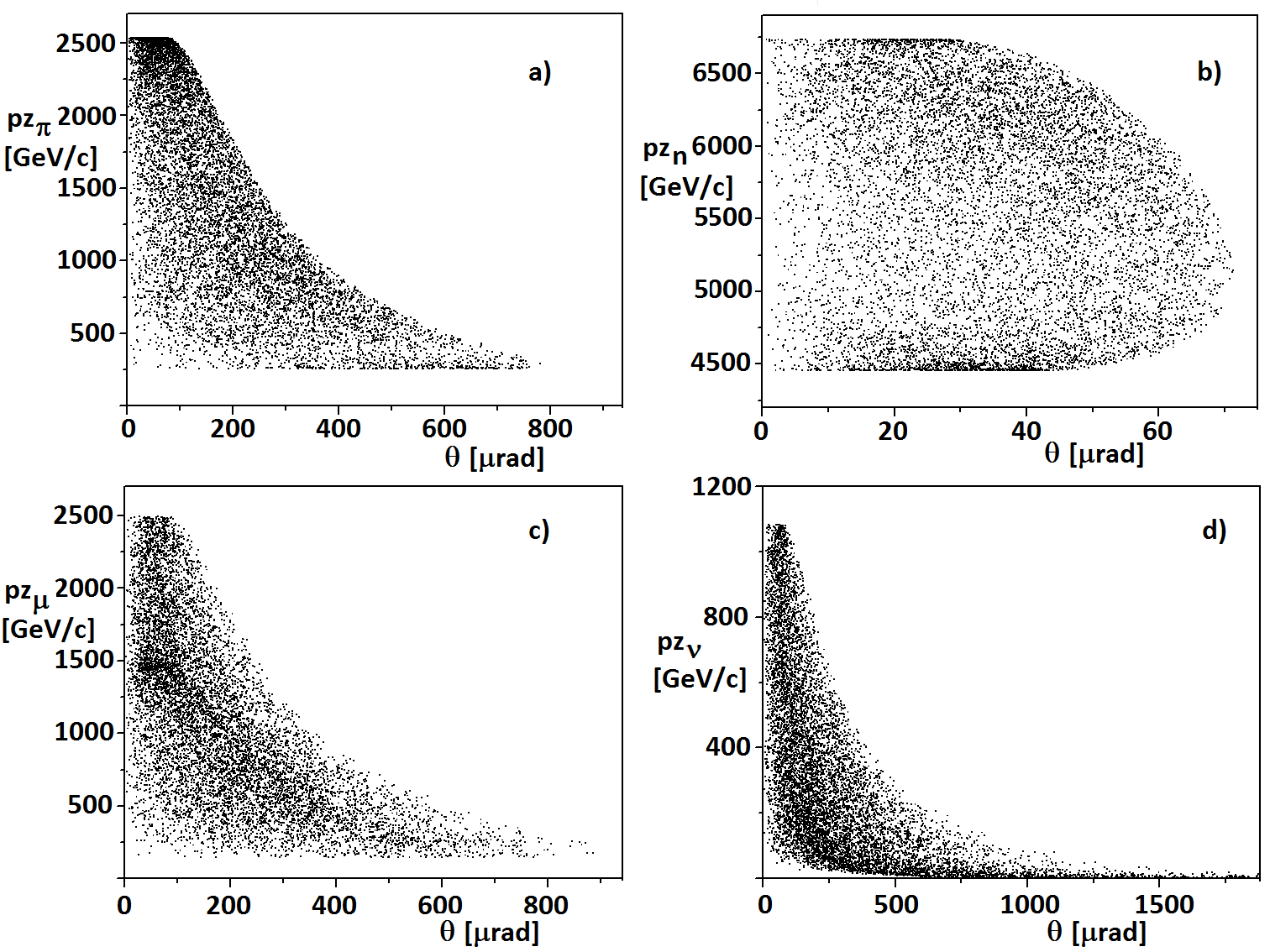}
	\caption{Pion, neutron, muon and neutrino longitudinal momentum [GeV/c] as a function of $\theta$ [$\mu$rad] (case $LHC$).}\label{part}
\end{figure}

\begin{figure}[htbp]
	\includegraphics [scale=0.21] {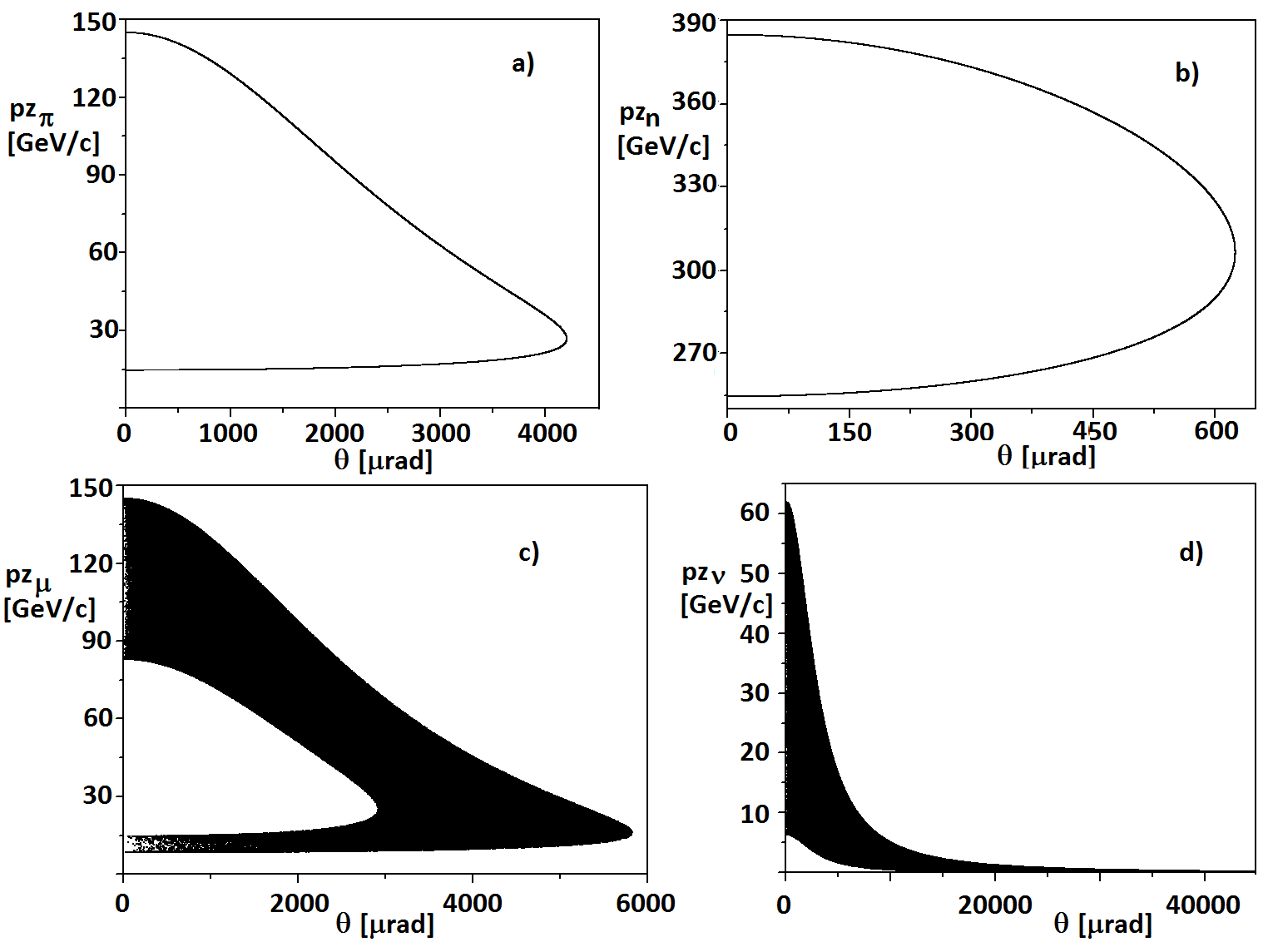}
	\caption{Pion, neutron, muon and neutrino longitudinal momentum [GeV/c] as a function of $\theta$ [$\mu$rad] (case $SPS$).}\label{part2}
\end{figure}

\begin{table}[htbp]
	\begin{tabular}{|c|c|c|c|c|}
		\hline  Particle & $pz_{min}$ & $pz_{max}$ & $px_{max}$ & $\theta_{rms}$ \\ 
		\hline  Proton $(pr)$ & $\simeq 7000$ & $7000$ & $\parallel beam$ & $\parallel beam$\\ 
		\hline  Photon $(ph)$ & $12$ keV & $12$ keV & $33$ meV & $1$ \\ 
		\hline  Pion $(\pi)$ & $520$ & $1509$ & $0.077$ & $26$ \\ 		
		\hline  Neutron $(n)$ & $5490$ & $6479$ & $0.077$ & $3.5$ \\  
		\hline  Muon $(\mu)$ & $298$ & $1509$ & $0.095$ & $31$ \\  
		\hline  Neutrino $(\nu)$ & $0$ & $644$ & $0.05$ &  $\le 500$  \\ 
		\hline 
	\end{tabular}
	\caption{Particles properties. Values of momenta in [GeV/c], angles in [$\mu$rad].}
	\label{tab:valorialtri2}
\end{table}
\begin{table}[htbp]
	\begin{tabular}{|c|c|c|c|c|}
		\hline  Particle & $pz_{min}$ & $pz_{max}$ & $px_{max}$ & $\theta_{rms}$ \\ 
		\hline  Proton $(pr)$ & $\simeq 50000$ & $50000$ & $\parallel beam$ & $\parallel beam$\\ 
		\hline  Photon $(ph)$ & $12$ keV & $12$ keV & $33$ meV & $1$ \\ 
		\hline  Pion $(\pi)$ & $407$ & $36431$ & $0.652$ & $36$ \\ 		
		\hline  Neutron $(n)$ & $13568$ & $49593$ & $0.652$ & $6.7$ \\  
		\hline  Muon $(\mu)$ & $233$ & $36431$ & $0.655$ & $37.6$ \\  
		\hline  Neutrino $(\nu)$ & $0$ & $15561$ & $0.281$ &  $\le 500$  \\ 
		\hline 
	\end{tabular}
	\caption{Particles properties. Values of momenta in [GeV/c], angles in [$\mu$rad].}
	\label{tab:valorialtri}
\end{table}
In graph $a)$ of Fig.\ref{pimu} and \ref{part2} the angular dependence of the longitudinal component of the pion momentum is shown. The more energetic branch comes from pions co-propagating with protons in $cm1$, the other one from the counter-propagating pions. The probability of a pion being produced at a certain angle in $cm1$ is proportional to the differential cross section (see \cite{drechsel}). In the $LHC$ case we also take into account the effect of the proton beam transverse emittance, which spreads the protons transverse momentum in phase space according to equation ${\sigma_{p_{x}pr}}\equiv M_{pr}\sqrt{\langle \overline{p}_{xpr}^{2}\rangle}=M_{pr}\epsilon_{xnpr}/\sigma_{xpr}$, where  ${\sigma_{p_{x}pr}}$ is the rms proton beam transverse momentum. The result of this operation is shown for all of the particles in Fig.\ref{part} while in Fig.\ref{pimu} is presented the comparison between the absence (graphs $a)$ and $b)$) and the presence (graphs $c)$ and d)) of the proton beam transverse emittance, which causes an enlargement of the angular spread and a dispersion in momentum both for primary and secondary particles. Considering $\epsilon_{xnpr}=1.4$ mm$\cdot$mrad as a typical value of $LHC$ proton beams and $\sigma_{xpr}=7$ $\mu$m we find $\sigma_{p_{x}pr}=187$ MeV/c ($\sigma_{x'pr}= 27$ $\mu$rad). 

It is interesting to notice that the phase space distributions shown in Fig.\ref{part} display a narrow stop-band at very small angles for all species of secondary particles much lighter than the proton. Pions, muons and neutrinos show a minimum angle of propagation in the laboratory reference frame of about $30$ $\mu$rad, close to the proton beam angular spread due to its emittance, while neutron phase spaces are clustered around this angle at the bottom and top energy bands (which correspond to neutrons propagating almost backward or almost forward in $cm1$).

As we see in Fig.\ref{pimu}, \ref{part} and \ref{part2} at the same angle in the laboratory frame we can find the same kind of particle at very different energies, therefore the selection of the particles has to be performed in the terms of energy. 

\begin{figure}[htbp]
	\includegraphics [scale=0.3] {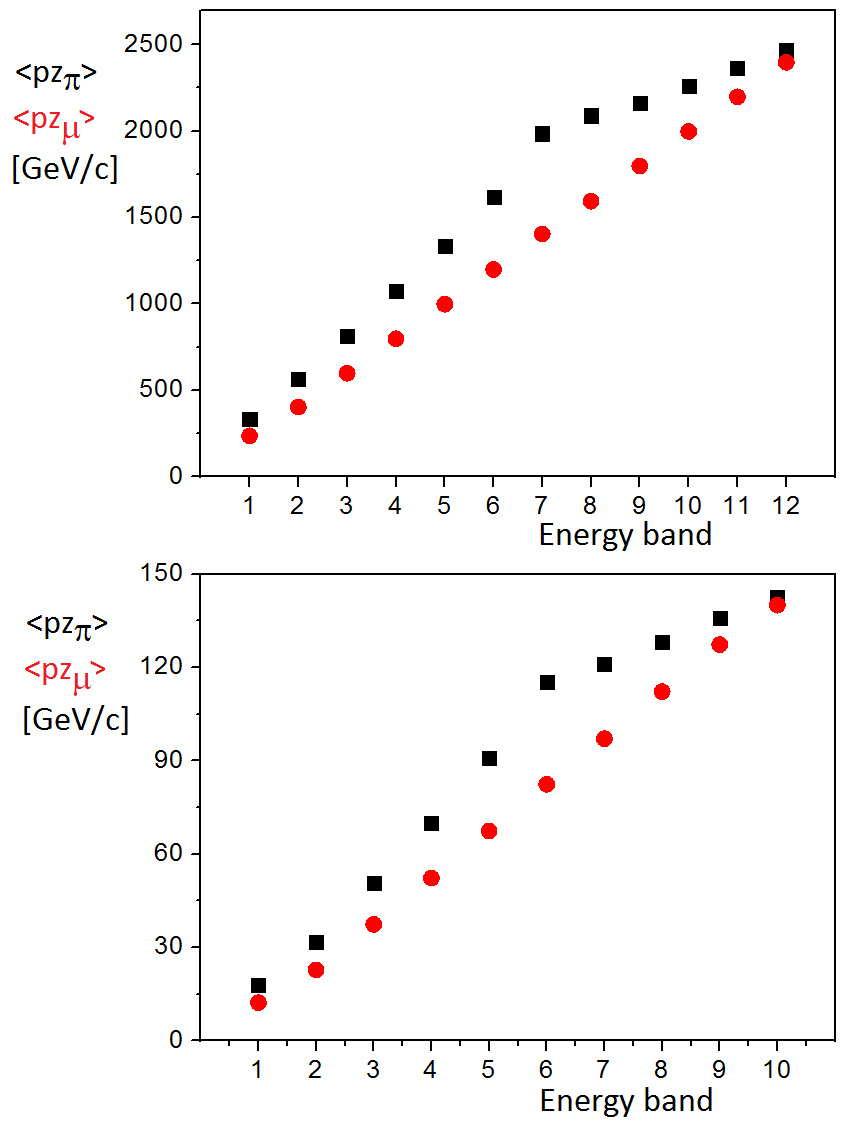}
	\caption{Mean value of the longitudinal momentum [GeV/c] of pions (black squares) and muons (red circles) as a function of the muon energy bands for $LHC$ (top) and $SPS$ (bottom) parameters. The energy bands range from the minimum to the maximum value of the muon longitudinal momentum,  with a $200$ (top) and $15$ (bottom) [GeV/c] binning.}\label{pzvari}
\end{figure}

The decay provides some special bands of longitudinal momentum of pions and muons which are almost overlapped, or restricted within a common narrow bandwidth, such that the muons in a possible pion storage ring would follow the same trajectory. As shown in Fig.\ref{pzvari} those bands turn out to be the less and the more energetic ones, where the mean value of pions and muons longitudinal momentum is almost coincident. 

\section{Luminosity and flux considerations}

Pions are produced at each collision between the proton bunch and the photon pulse according to the collider luminosity, given by
\begin{equation}
	L_{prph} = \frac{N_{pr}N_{ph}r}{4\pi\sigma_{xpr}^2} 
\label{lum}\end{equation}
where $N_{pr}$ is the number of colliding protons in the bunch, $N_{ph}$ is the number of photons carried by the radiation pulse, $r$ the repetition rate of the collisions and $\sigma_{xpr}$ the  effective spot size at the IP. Since the interaction does not produce beam-beam effects, due to the nature of the photon pulse and since the collective e.m. field of the X-FEL photon beam is very small (the equivalent laser parameter representing the dimensionless value of its e.m. vector potential as seen by the proton $a_{0}\propto10^{-4}(M_{e}/M_{pr})\lambda_{FEL}[\text{\AA}]\sqrt{P[TW]}/\sigma_{xh\nu}[\mu m]$ is much smaller than 1), we can assume that the geometrical emittance given by eq.(\ref{lum}) is a good assumption as far as the hour-glass effects are negligible. This is certainly the case for an optimized $LHC$ or $FCC$ beam at the IP (with a minimum beta function of about $0.1$ m, matched to the proton bunch length to mitigate the hour-glass effect), and it is even more valid for the XFEL beam, since its geometrical emittance is much smaller than the proton beam one ($\epsilon_{xpr}=\epsilon_{xnpr}/\gamma_{pr}=2 \cdot 10^{-10}$ m$\cdot$rad for the proton beam at $7$ TeV and $\epsilon_{xh\nu}=8\cdot 10^{-12}$ m$\cdot$rad for the XFEL photon beam). We should also remark that the FEL photon pulse is extremely short ($\simeq 200$ fs) compared to the proton bunch, with a diffraction Rayleigh range $Z_{0}=2\pi\sigma_{h\nu}^{2}/\lambda_{FEL}$ longer the $1$ m: hour-glass effects are therefore depending only on the proton beam diffraction at IP.

Concerning the maximum repetition rate achievable in collisions, $LHC$ can be operated up to $40$ MHz and $FCC$ is expected to match it, while XFEL based on CW SC Linacs are aimed at achieving $1$ MHz ($LCLS-II$), but the upgrade to $10$ MHz is considered feasible by adopting $ERL$ schemes, which are definitely much more expensive and challenging but performant in terms of electron beam average current (for our case we would need an average current of about $0.2$ nC $\times$ $10$ MHz $=2$ mA, which is certainly in the range of capability of an $ERL$).

Eventually, considering ultimate performances for either $LHC/FCC$ beams and XFELs beams, we adopt these values for the evaluation of fluxes and proton beam life-time due to event production: $N_{ph}=2\cdot 10^{14}$, $N_{pr}=2\cdot 10^{12}$, $r=10$ MHz and $\sigma_{xpr}=7$ $\mu$m. This implies $L_{prph}=6.5\cdot 10^{38}$ cm$^{-2}$s$^{-1}$. So the total number of pions $N_{\pi}$ produced per bunch crossing is about $1.6\cdot 10^{4}$ assuming a maximum of the total cross section $\Sigma_{\pi}$ of about $220$ $\mu$barn ($1.6\cdot 10^{11}$ pions/s average flux), as given by the formula  $N_{\pi}=L_{prph}\Sigma_{\pi}$. There are small differences in the total cross section for the following reactions:
\begin{equation}\begin{split}
	p+h\nu\rightarrow\pi^{+}+n \hspace{1cm} n+h\nu\rightarrow\pi^{-}+p\\ p+h\nu\rightarrow\pi^{0}+p  \hspace{1cm} n+h\nu\rightarrow\pi^{0}+n
\end{split}
\end{equation}
so we should expect an almost equivalent production of charged and neutral pions. These latter are not of interest for a muon collider application since they decay very rapidly (mean life-time $8.4\cdot 10^{-17}$ s) into a pair of high energy $\gamma$ photons.

Clearly for a muon-antimuon collider one needs to produce antimuons via the reaction $n+h\nu\rightarrow\pi^{-}+p$, followed by the decay of $\pi^{-}$ into a $\mu^{-}$ and a neutrino. That implies to collide the FEL photon beam against a deuteron TeV beam, exploiting the symmetry in positively and negatively charged pion photo-production on deuteron (see \cite{darwish}). For this reason we analyze the ultimate case of $FCC$ proton beams considering $50$ TeV energy per nucleon to take into account the option of accelerating deuterons at $50$ TeV/n (nominal $100$ TeV for protons) and colliding this beam with the XFEL photon beam.

\begin{figure}[htbp]
	\includegraphics [scale=0.3] {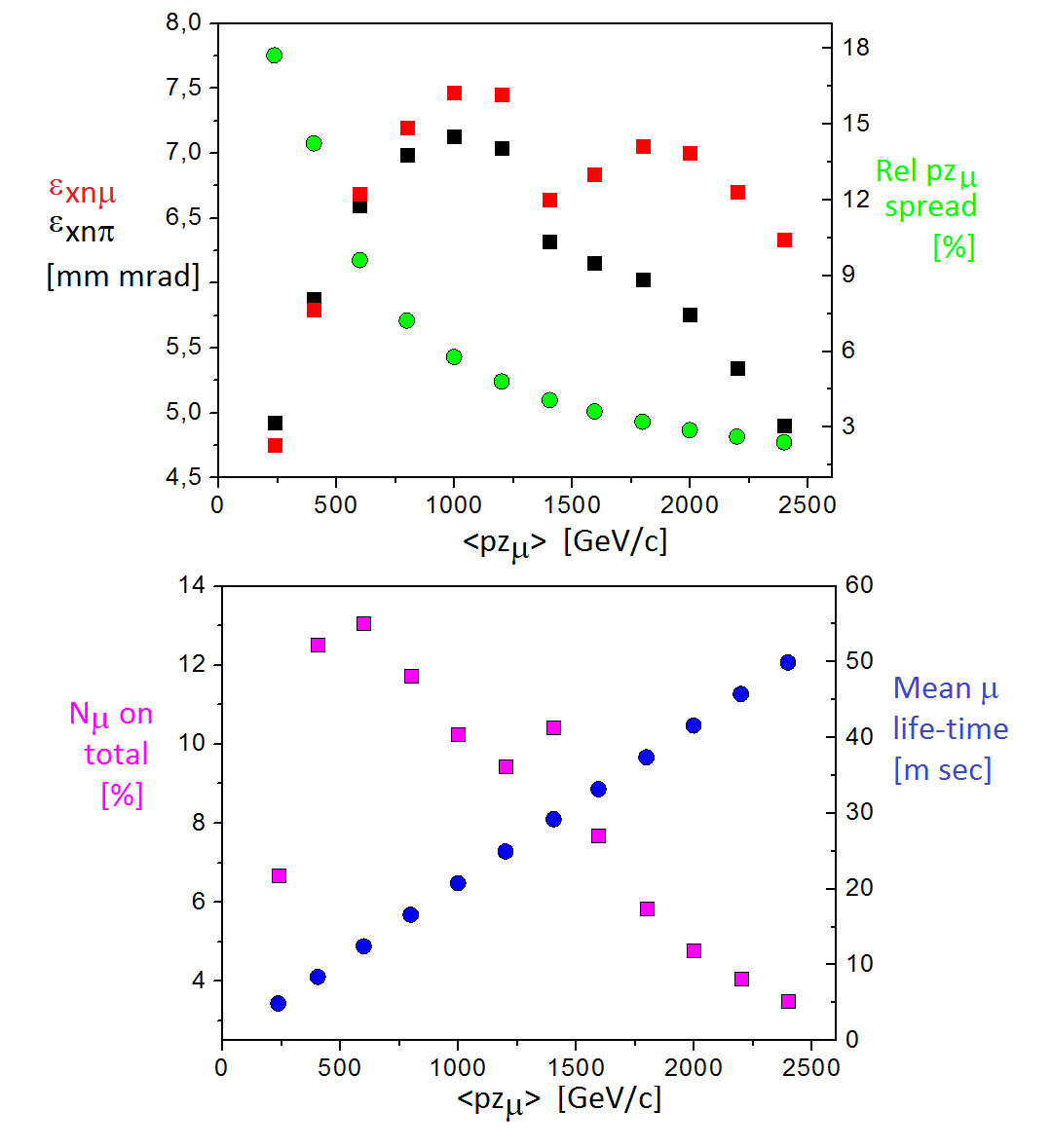}
	\caption{Muons properties (case $LHC$): muon normalized emittance [mm$\cdot$mrad] in red, pion normalized emittance [mm$\cdot$mrad] in black, relative longitudinal momentum spread [$\%$] in green, relative number of muon on total [$\%$] in magenta and muon mean life-time [ms] in blue as a function of the mean longitudinal momentum of the muons [GeV/c].}\label{tabe}
\end{figure}

\begin{figure}[htbp]
	\includegraphics [scale=0.3] {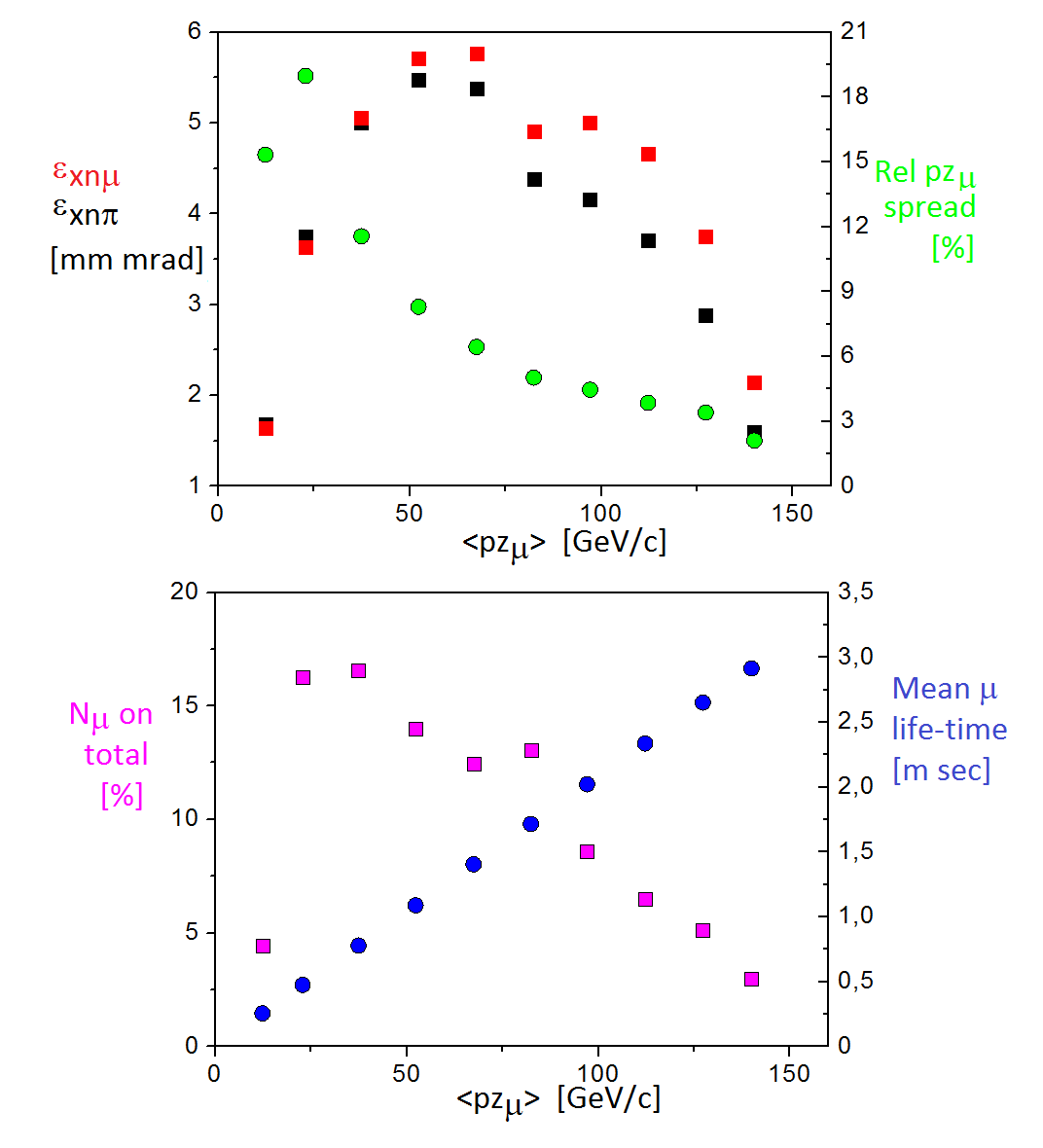}
	\caption{Muons properties (case $SPS$): muon normalized emittance [mm$\cdot$mrad] in red, pion normalized emittance [mm$\cdot$mrad] in black, relative longitudinal momentum spread [$\%$] in green, relative number of muon on total [$\%$] in magenta and muon mean life-time [ms] in blue as a function of the mean longitudinal momentum of the muons [GeV/c].}\label{tabe2}
\end{figure}

Under the assumption of preservation of the outstanding pion beam emittance, and considering the long life-times of muons at the TeV level (in excess of tens of milliseconds), one can conceive a storage ring accumulating muons, in order to go from the small number of pions produced at each bunch crossing (about $10^{4}$) up to the requested muons per bunch typical of a muon collider. Assuming this latter value and a life-time of $20$ ms for the muons (see Fig.\ref{tabe} and \ref{tabe2}) we would accumulate $10^{9}$ muons during half of their life-time, which in turns is enough to achieve an effective muon collider luminosity of about $10^{31}$ s$^{-1}$cm$^{-2}$ at a few $\mu$m IP spot size, interleaving in time accumulation and collision of the muon bunches at each muon half life-time.
As shown in Fig.\ref{tabe} and \ref{tabe2} the concentration of muons has a maximum at low energy and a secondary peak at a mean energy as we can notice in $LCH$ case also from graph $c)$ in Fig.\ref{part}.  \\

There are other reactions possible in the collision between the XFEL photons and the relativistic proton/deuteron beam, namely the Compton back-scattering of the photons by the proton, the electron/positron pair production in the nuclear field of the deuteron and the direct muon pair photo-production (so called Bethe-Heitler lepton-pair production process, $p+h\nu\rightarrow p'+ e^{+} e^{-}$, $p+h\nu\rightarrow p'+ \mu^{+} \mu^{-}$).
The Compton scattering is characterized by a total cross section scaling like $\Sigma_{T}(M_{e}/M_{pr})^{2}$ , so it lays in the range of a fraction of $\mu$barn, quite negligible both for secondary beam generation and for the budget of events setting the life-time of the deuteron beam ($\Sigma_{T}$ is the Thomson cross section, about $0.67$ barn \cite{jackson, vittoria}).

The electron/positron pair production in the field of the nucleon is instead the dominant process in the collision deuteron-photon, since its total cross section  $\Sigma_{e^{+}e^{-}}$ in the range of a few hundreds MeV photons (in the rest frame of the nucleon) is about $15$ mbarn (see \cite{fujii}). So we expect a number of electron/positron pair created at any bunch crossing of the order of $1.1\cdot 10^{6}$. As discussed in the previous chapter, these pairs are propagating as an intense collimated secondary beam along with the pion beam.

The direct photo-production of muon pairs in the nucleon field by the FEL photon is on the other side characterized by a very small total cross section, which indeed scales like $\Sigma_{e^{+}e^{-}}(M_{e}/M_{\mu})^{2}$, so it is smaller than a $\mu$barn \cite{mcdonald}, therefore yielding about $2.3 \cdot 10^{8}$ muon pairs per second at $10$ MHz repetition rate.

The deuteron beam life-time can be evaluated just rescaling $LHC$ data (beam life-time of about $40$ hours) for $p-p$ events, which are characterized by a total cross section of about $110$ mbarn and a luminosity (present operation) of $10^{34}$ s$^{-1}$cm$^{-2}$: running the deuteron-photon collider at a luminosity of $10^{38}$ s$^{-1}$cm$^{-2}$ we would get, at a dominant cross section of $15$ mbarn, a beam life-time of the order of a second, which sets up a serious challenge to the injection chain of the main ring (a top-up injection scheme is likely to be implemented).

\begin{table}[htbp]
	\begin{tabular}{|c|c|c|c|}
		\hline  $e^{+}/e^{-}$ & $pz_{min}$ & $pz_{max}$ & $px_{max}$ \\ 
		\hline  $LHC$ & $0.006$ & $1362$ & $0.116$ \\ 
		\hline  $FCC$ & $0.043$ & $10136$ & $0.123$ \\ 
		\hline 
	\end{tabular}
	\caption{$e^{+}/e^{-}$ properties (cases $LCH$ and $FCC$). Values of momenta in [GeV/c].}
	\label{tab:valoricoppie}
\end{table}

We did not evaluate the cascaded interaction of the electron/positron pair with the counter-propagating FEL colliding photon beam and we did not carried out the generation of its corresponding phase space distributions. This cascaded interaction is indeed possible due to the highly collimated behavior of secondary beams in this highly Lorentz boosted interaction, unlike p-p interactions at $LHC$ IP which occur with a steady c.m. reference frame. Since part of the $10^{6}$ electron/positron pairs will propagate after being generated along with the surviving proton beam, they may collide with a fraction of the incoming $10^{14}$ FEL photons giving rise to muon pair production via the reactions
\begin{equation}
e^{-} + h\nu \rightarrow  e^{-} + \mu^{+}\mu^{-}  \hspace{1cm} e^{+} + h\nu \rightarrow  e^{+} + \mu^{+}\mu^{-}  
\end{equation}
As a rough evaluation of the total number of muon pairs produced per bunch crossing we take as an example $2$ TeV electrons or positrons created in the proton-photon collision (nearly $10^{6}$ per bunch crossing), and we consider their cascaded collision with FEL photons at $12$ keV. The energy available in the center of mass is about $309$ MeV, well above the threshold of $2 M_{\mu}$. Following Ref. \cite{athar} we use a total cross section for the muon pair photo-production of about $30$ nbarn, which gives an upper limit for the muon pairs produced of about $5$ per second (although collimated within an incredibly small forward angle of about $0.3$ $\mu$rad). Just as a side comment we notice that the direct muon pair photo-production in our primary proton-photon collision via Bethe-Heitler process can give a flux of about a few $10^{8}$ muon pairs per second as discussed above in this section.

\section{conclusions}

The combination of two most advanced accelerators, the Large Hadron Collider and the X-ray Free Electron Laser, although looking as a strange and unmatched marriage, and definitely not easy nor inexpensive to be implemented in reality, offers the great opportunity of conceiving an hybrid proton-photon collider, the Hadron Photon Collider ($HPC$), at an unprecedented luminosity exceeding $10^{38}$ s$^{-1}$cm$^{-2}$. The $HPC$ is actually aimed not at producing events to study, but to generate secondary beams of unique characteristics, via a highly boosted Lorentz frame corresponding to a very relativistic moving center of mass reference frame. The opportunity may become quite strategic in the frame of a new machine under design and being proposed like Future Circular Collider, since the higher energy foreseen ($100$ TeV) greatly simplifies the type of X-ray FEL machine that should be coupled for implementing a $HPC$: in fact at $FCC$ the optimal FEL photon energy is about $3$ keV ($\lambda_{FEL}=0.4 
$ nm), easily achievable by a CW SC GeV-class Linac driving the FEL radiator. 

As shown in sections II and III, the phase space distributions of the secondary beam generated, in particular pions (charged and neutral), have outstanding properties of low transverse emittance (in the range $4-8$ mm$\cdot$mrad) and are collimated within very narrow forward angles (less than $150 $ $\mu$rad for a $7 $ TeV colliding proton beam as for $LHC$ and less $10$ $\mu$rad for a $50$ TeV/n deuteron beam as for $FCC$) with energies attaining $2.5$ TeV and $19.1$ TeV respectively for the two cases ($LHC$ and $FCC$). 

The opportunity consists in conceiving a beam manipulation of charged pions of both signs, generated in the $HPC$ on deuterons, capable to let them decay into muons and neutrino beams of similar characteristics in the phase space: the challenge is the large energy spread of the pion beams (an energy selection will have to be implemented first) and the long range distance travelled by them before decay, at this energy. The potentialities in terms of muon and neutrino beams obtainable are impressive: although the number of muons and neutrinos per bunch would be low (a few thousands) their phase space distributions would be extremely high quality thanks to the large Lorentz boost of the primary proton-photon collision, giving rise to very small divergence angles for these beams (tens of $\mu$rads for muons and hundreds of $\mu$rads for neutrinos). The long life of the muons generated by decaying pions (in excess of $10$ ms) may offer the opportunity to accumulate them in a storage ring so to achieve muon collider requested bunch intensities. 

TeV bunches of up to $10^{6}$ of electron positron pairs are also generated with very good emittances and collimation properties, while direct production of muon pairs in the primary proton-photon collision at $HPC$ is quite low efficient (a few tens per bunch crossing) but perhaps of some interest due the highly collimated characteristic of the muon pair beam.

\acknowledgments
We acknowledge very fruitful conversations and quite inspirational feedbacks from C. Meroni, A. Ghigo, D. Palmer, L. Rossi, R. Garoby, D. Babusci and F. Zimmermann.

\end{document}